\documentclass[reprint, nofootinbib, amsmath, amssymb, aps, prx, superscriptaddress]{revtex4-2}
\usepackage{graphicx} 
\usepackage{braket} 
\usepackage{bm}
\usepackage{booktabs}
\usepackage{array}
\newcolumntype{C}[1]{>{\centering\arraybackslash}p{#1}}
\newcolumntype{L}[1]{>{\raggedright\arraybackslash}p{#1}}
\usepackage{tikz}
\usepackage{pgfplots}
\usepackage{xcolor}
\usepackage{colortbl}
\usepackage{bbm}
\usepackage{amsfonts}
\usepackage{subcaption}
\usepackage{mathtools}
\usepackage{tabularx}   
\usepackage{makecell} 
\usepackage{LVcolours} 
\usepackage{appendix}
\usepackage{algpseudocode}

\makeatletter
\newcommand\NoMarkerFootnote[1]{%
  \begingroup
  \renewcommand\@makefnmark{}%
  \footnotetext{#1}%
  \endgroup
}
\makeatother

\usepackage[hidelinks, colorlinks=true]{hyperref}

\begin{document}

\title{Hybrid Quantum-Classical Learning for Multiclass Image Classification}

\author{Shuchismita Anwar}
\affiliation{Department of Computer Science and Engineering, Brac University, Dhaka 1212, Bangladesh}

\author{Sowmitra Das}
\affiliation{Department of Computer Science and Engineering, Brac University, Dhaka 1212, Bangladesh}
\affiliation{Blackett Laboratory, Imperial College London, London SW7 2AZ, United Kingdom}

\author{Muhammad Iqbal Hossain}
\affiliation{Department of Computer Science and Engineering, Brac University, Dhaka 1212, Bangladesh}

\author{Jishnu Mahmud}
\affiliation{Department of Computer Science and Engineering, Brac University, Dhaka 1212, Bangladesh}


\begin{abstract}
This study explores the challenge of improving multiclass image classification through quantum machine‑learning techniques. It explores how the discarded qubit states of Noisy Intermediate‑Scale Quantum (NISQ) quantum convolutional neural networks (QCNNs) can be leveraged alongside a classical classifier to improve classification performance. Current QCNNs discard qubit states after pooling; yet, unlike classical pooling, these qubits often remain entangled with the retained ones, meaning valuable correlated information is lost. We experiment with recycling this information and combining it with the conventional measurements from the retained qubits. Accordingly, we propose a hybrid quantum–classical architecture that couples a modified QCNN with fully connected classical layers. Two shallow fully connected (FC) heads separately process measurements from retained and discarded qubits, whose outputs are ensembled before a final classification layer. Joint optimisation with a classical cross‑entropy loss allows both quantum and classical parameters to adapt coherently. The method outperforms comparable lightweight models on MNIST, Fashion‑MNIST and OrganAMNIST. These results indicate that reusing discarded qubit information is a promising approach for future hybrid quantum–classical models and may extend to tasks beyond image classification.
\end{abstract}


\maketitle
\NoMarkerFootnote{\textit{This is a preprint.}}

\section{Introduction}\label{sec1}
Quantum machine learning combines the representational power of artificial intelligence with the parallelism and entanglement of quantum computing, offering a practical route to tackle computationally intensive tasks such as multiclass image recognition \cite{henderson2019quanvolutional, s23052753, ciliberto2018quantum}. While classical convolutional neural networks (CNNs) have long dominated visual recognition, their accuracy gains come at the cost of ever-growing depth, parameter counts, and energy consumption \cite{kim2023classical, Li_2020, vatan2004optimal, Chalumuri2}. Quantum Convolutional Neural Networks (QCNNs) mitigate these overheads by executing convolution and pooling through unitary operations, achieving comparable or occasionally superior performance with far fewer parameters and shallower circuits, an advantage that is especially relevant for today’s noisy intermediate-scale quantum (NISQ) hardware \cite{henderson2019quanvolutional, kim2023classical, Cong2019, angle2, liang2021hybrid, Preskill_2018, Bharti_2022, Philip, Hur2022, Chandra, Wei2022, Abbas}.

Empirical studies already report strong QCNN accuracies on benchmarks such as MNIST, Fashion-MNIST, and various medical-imaging datasets \cite{henderson2019quanvolutional, kim2023classical, Philip, Hur2022, Chandra}. For example, Henderson et al.\ surpassed 88\% on a brain-disease diagnostic task despite using only tens of quantum parameters \cite{bokhan2022multiclass}. Hybrid quantum–classical models (HQCNNs) extend this promise, delivering competitive results on tabular problems like Iris and on Wi-Fi-based indoor localisation, where the quantum front-end extracts low-dimensional yet highly separable features \cite{Hur2022, Chalumuri}.

Recent work has moved beyond raw circuit design toward specialised quantum embeddings, lightweight unitary ansätze, and problem-specific classifiers, often outperforming classical networks with an order of magnitude fewer parameters \cite{s23052753, Li_2020, Chalumuri2, angle2, Chandra, Chalumuri, lloyd2020quantum, Jain, Kunkun, angle1}. These gains are frequently attributed to embeddings that maximise class separation in Hilbert space, thereby reducing the burden on the subsequent classifier \cite{s23052753, angle2, Chandra, Wei2022, Chalumuri, angle1}. Parallel advances in Quantum Support Vector Machines highlight the same trend: carefully engineered quantum feature maps can deliver superior decision boundaries with logarithmic complexity in the data dimension \cite{Rebentrost_2014, lloyd2014quantum}.

Despite promising results; 98.97\% on MNIST and 91.40\% on GTSRB for Quantum Deep CNNs \cite{Li_2020}, for example - scale and noise resilience remain open challenges as datasets grow in size and complexity \cite{ciliberto2018quantum, Philip, Shiqin, Liu_2021, larose2020robust, zhao2019smoothinputpreparationquantum, farhi2018classificationquantumneuralnetworks, Schuld_2020, Haah_2019, kerenidis2016quantumrecommendationsystems, Kerenidis_2020, dervovic2018quantumlinearsystemsalgorithms}. Novel variants such as the Amplitude-Transformed QCNN (ATQCNN) improve data-efficiency for domain-specific inputs, yet still struggle on heterogeneous datasets when the parameter budget is tightly capped \cite{Shiqin}. Tensor-network methods achieve above-98\% accuracy in binary settings but often falter on multiclass tasks due to limited expressive power once the bond dimension is fixed \cite{Kunkun}. Other hybrid schemes have demonstrated utility in medical imaging; for example, Liang et al.\ for Alzheimer’s detection and Lusnig et al.\ for federated liver analysis yet they too inherit the same bottleneck as they scale beyond binary or small-class problems \cite{liang2021hybrid, lusnig2023hybrid, Emine}.

Traditional and hybrid QCNNs achieve dimensionality reduction through pooling layers by omitting qubits, which causes quantum information loss and compromised performance, especially in NISQ devices with scarce qubits \cite{Wei2022, Philip, lusnig2023hybrid, angle1, garcia2021quantum}. However, previous research has not sufficiently addressed the limitations arising from discarded quantum information during pooling, nor has it explored fully exploiting quantum-classical interactions to enhance classification performance. This gap significantly limits QCNN potential for complex real-world data applications.

To address these challenges, in this work, we suggest a novel hybrid quantum-classical architecture with improved classification performance, conserving quantum information, and increased resource efficiency. The following work proposes a novel Hybrid Quantum-Classical Architecture that addresses a number of challenges with several important contributions:

\begin{itemize}
    \item \emph{Proposed Scaling Method}: The architecture proposes to extend the range of quantum measurement values which optimally maps quantum data onto the classical values. The scaling of the measurements ensures more precise and efficient quantum operations which raises the accuracy of the hybrid model during the classification process. 

    \item \emph{Utilization of Discarded Qubits from the Pooling Layer}: Instead of discarding the qubit states completely from the pooling layer like traditional QCNN's, the proposed architecture reuses such discarded qubits. Further processing extracts useful quantum information from them; it also adds it to the final computations and thereby improving the performance in classification with limited computational overhead. The method avoids loss of information as well as strengthens the capability of the model to handle complicated data.
    
    \item \emph{Classical Layers for Joint Optimization}: The scaled values of discarded qubits and the retained qubits undergo classical layers separately. After that, their outputs get dot-multiplied. This captures more of the quantum features that will contribute in the overall joint optimization of the model. These classical layers back-propagate to dynamically adjust parameters on the combined quantum-classical data. The classical layers further optimize decision boundaries for improved classification results that combine the strengths of both quantum and classical worlds.
    
\end{itemize}

The remainder of this article is arranged as follows.  
Section~\ref{Proposed Architecture} motivates and describes the full hybrid architecture, outlining its three principal subsystems-quantum encoding, quantum convolution as well as pooling, and classical post‐processing.  
Section~\ref{Embedding Techniques} \& Section~\ref{Quantum Circuit Components} drills down into the circuit‐level mechanisms, covering the chosen embeddings, the SU(4) convolutional template, the pooling mechanism, and the proposed methods. Section~\ref{Evaluation of Proposed Hybrid Quantum-Classical Model} reports quantitative results, including ablation curves and head‐to‐head comparisons with prior lightweight QCNNs. Finally, Section~\ref{Conclusion} concludes the key findings, discusses current limitations, and sketches directions for future research.  An appendix provides complete implementation details code versions, hardware configurations, and reproducibility scripts to facilitate independent verification and extension of our work.

Unlike prior accuracy-boosting tricks that rely on deeper circuits, extra ancilla qubits, or thousands of classical weights \cite{henderson2019quanvolutional,Li_2020}, our method simply re-uses qubits that the QCNN already discards and feeds them through a single shallow dense layer. Thus the quantum register size, gate depth, and noise budget stay unchanged while test accuracy rises up on four-class MNIST, Fashion-MNIST, and OrganAMNIST.  In short, the scheme delivers better performance with zero additional quantum resources and a negligible classical footprint.

We acknowledge that most existing studies, including the present one evaluate QCNNs on classical datasets because public image benchmarks provide an unambiguous yard-stick for measuring incremental architectural gains (e.g.\ our discarded-qubit reuse).  
Nevertheless, QCNNs were originally conceived for quantum‐state discrimination, such as identifying quantum phases or error-syndrome patterns~\cite{Cong2019}. Although our benchmarks focus on classical image data, the hybrid QCNN is independent of how the input state is prepared. Extending the evaluation to intrinsically quantum problems, such as topological-phase recognition or error syndrome classification, remains a natural direction for future work.

\section{Proposed Architecture}\label{Proposed Architecture}
\begin{figure*}[!ht]
    \centering
    \hspace*{-1.05cm}
    \includegraphics[width=1\textwidth]{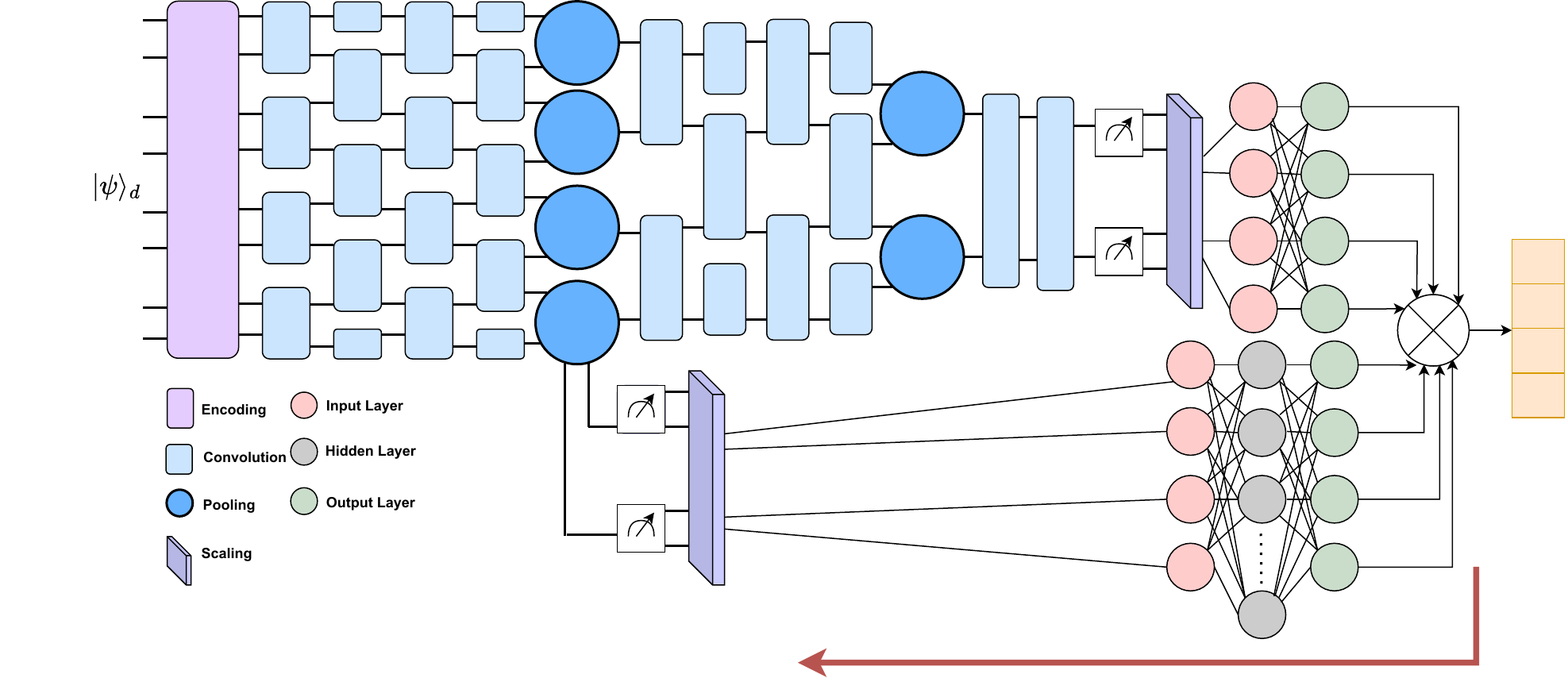}
    \caption[Proposed Hybrid Quantum-Classical Convolutional Neural Network (QCNN)]{
        The architecture includes a Hybrid Quantum-Classical Convolutional Neural Network (QCNN) containing three major quantum layers that process the input data: encoding, convolution, and pooling. Both the retained and discarded qubits feed into the classical fully-connected layers to perform classification. In this architecture, quantum feature extraction is combined with a classifier.
    }
    \label{fig:model}
\end{figure*}

The proposed hybrid quantum–classical architecture embeds each pre-processed image into a quantum state; using amplitude or angle encoding, selected automatically according to input dimensionality and channels the register through three repeating blocks (Fig.~\ref{fig:model}). Pooling halves the qubit count but instead of discarding the measured qubits outright, the design captures the probabilities of both the retained and the discarded qubits, then linearly rescales them to the symmetric interval $[-2,2]$ to stabilise classical optimisation. Each probability vector flows into a dedicated shallow fully connected (FC) head that refines the features; the two refined vectors are fused by element-wise Hadamard product and passed to a final FC layer that outputs the class logits. Recycling qubits that would otherwise be lost preserves entanglement-mediated correlations, the shared affine rescaling layer improves gradient flow without adding trainable parameters, and end-to-end training lets quantum gates and classical weights co-adapt under a single cross-entropy loss. Together these elements raise test accuracy on MNIST, Fashion-MNIST, and OrganAMNIST while keeping circuit depth and parameter count compatible with current NISQ devices.

\section{Embedding Techniques}\label{Embedding Techniques}

The hybrid pipeline begins by mapping pre-processed pixel vectors onto quantum
states in an optimized resource utilization \cite{lloyd2014quantum,Rebentrost_2014,araujo2021divide}. Two complementary encoders are employed, each tailored to a different feature-dimensionality range.

\subsection{Amplitude embedding}\label{ssec:amp}

When the feature vector is high-dimensional, we adopt \textit{amplitude embedding}.  
Let $\boldsymbol{x}=(x_0,\dots,x_{N-1})^{\mathsf T}$ be the normalised input such that $\sum_{i=0}^{N-1}|x_i|^{2}=1$, where each $x_i$ is the scalar amplitude associated with basis state $\ket{i}$.  
For $N=2^{n}$ features we prepare an $n$-qubit register in  
\begin{equation}
  \ket{\psi(\boldsymbol{x})}
  \;=\;
  \sum_{i=0}^{N-1} x_i\,\ket{i}.
\end{equation}
No additional gates are needed once this state is loaded, leaving the full expressive power of the amplitudes available to the subsequent convolutional layers.  
In our experiments we set $n=8$, allowing up to $2^{8}=256$ features while remaining within NISQ hardware limits \cite{Shiqin}.

\subsection{Angle embedding}\label{ssec:angle}

For lower-dimensional inputs we prefer \emph{angle embedding}, which trades
amplitude capacity for shallower depth and improved noise tolerance.  Each
feature is translated into a pair of single-qubit rotations,
\begin{align}
  R_Y(\theta_i) &= e^{-i\theta_i\sigma_y/2}, &
  R_Z(\phi_i)  &= e^{-i\phi_i\sigma_z/2},
\end{align}
giving the product state
\[
  \ket{\psi(\boldsymbol{x})}
  =
  \bigotimes_{i=0}^{n-1} R_Z(\phi_i)\,R_Y(\theta_i)\,\ket{0},
\]
where the angles $(\theta_i,\phi_i)$ are affine maps of the normalised pixel intensities, ensuring uniform coverage of the Bloch sphere \cite{angle1,angle2,larose2020robust}.  
Because each data dimension consumes exactly one qubit, the encoder scales linearly with hardware resources while keeping the gate count minimal.  
All reported results use $n=8$ qubits for a fair comparison with the amplitude-embedding configuration.
After min–max normalisation $x_i\!\in\![0,1]$, each pixel is converted to a
pair of rotation angles by an affine map
\[
   \theta_i = \pi\,x_i,
   \qquad
   \phi_i   = \pi\,(2x_i-1).
\]
The factor $\pi$ ensures full coverage of the Bloch sphere; the
$\bigl(2x_i-1\bigr)$ term centres the $R_Z$ phase around~$0$ so that
$\phi_i\!\in\![-\pi,\pi]$ while $\theta_i\!\in\![0,\pi]$.

\paragraph{Choice of encoder}
Amplitude embedding is selected whenever the pre-processing pipeline outputs a
vector with $N\!>\!8$ elements; otherwise, angle embedding is applied.  This
adaptive strategy balances qubit usage against circuit depth, providing the
quantum backbone with an information-rich starting point while remaining
compatible.

\begin{figure*}[!ht]
    \centering
    \includegraphics[width=0.65\textwidth]{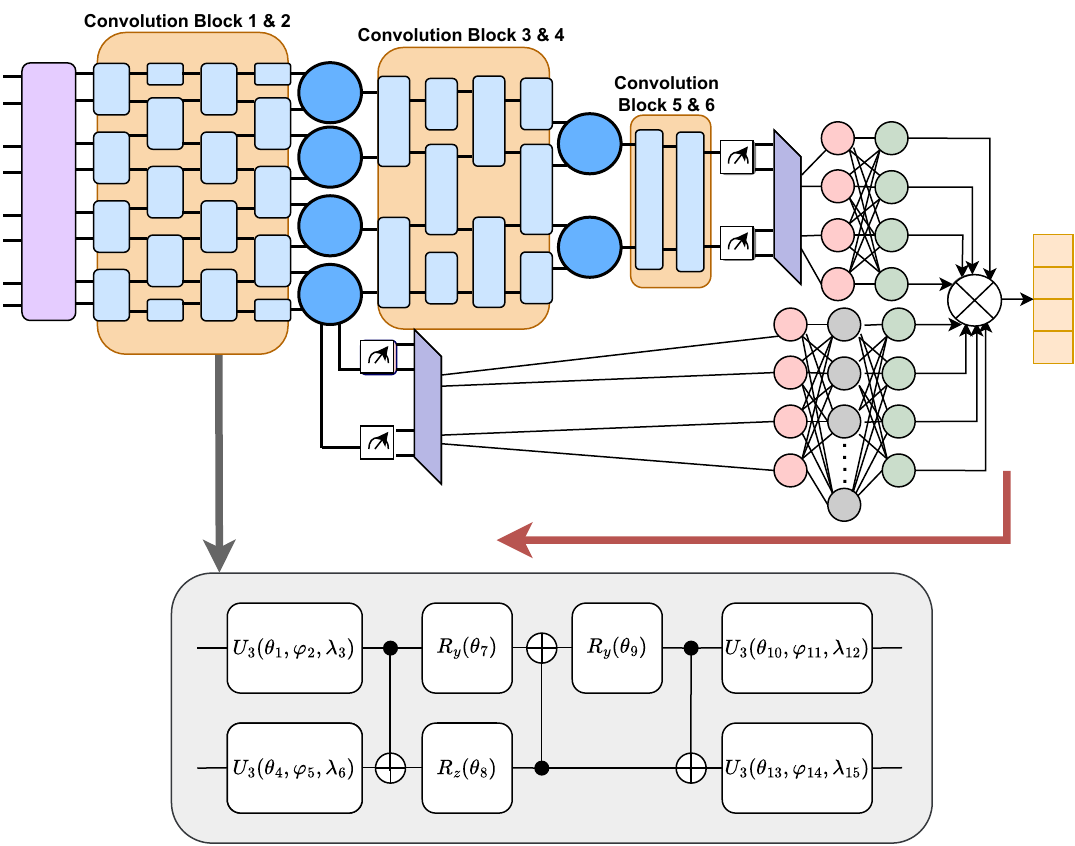}
    \caption[Quantum Convolution Layer Structure]{
        Quantum Convolution Layer Structure. This figure represents the quantum convolutional layers with the \( U_{SU4} \) ansatz, having 15 parameters, \( \theta, \phi, \lambda \) respectively. Different gates, such as \( U_3 \), CNOT, \( RZ \), and \( RX \), have been applied here to extract features from the quantum state in order to gather patterns and develop high-level abstract representations progressively.
    }
    \label{fig:conv}
\end{figure*}

\section{Quantum Circuit Components}\label{Quantum Circuit Components}

\subsection{Quantum Convolutional Module}

Any two-qubit operation can be expressed as an element of \(\mathrm{SU}(4)\), the set of all \(4\times4\) unitary matrices whose determinant equals one.  Because a complex \(4\times4\) matrix has \(4^{2}=16\) complex entries and the unitary plus determinant constraints together remove one complex and one real degree of freedom, a general member of \(\mathrm{SU}(4)\) is specified by \(4^{2}-1 = 15\) independent real parameters \cite{Ragone2024}.  \textit{This one-to-one match between parameters and degrees of freedom guarantees that our ansatz is universal for two-qubit gates while avoiding redundant trainable variables, enabling shallow yet expressive layers that compile cleanly on hardware with nearest-neighbour connectivity.}  Each convolutional layer in our network therefore applies \(l_i\) copies of a 15-parameter template \(U_{\text{conv}}(\boldsymbol{\theta})\) to nearest-neighbour qubit pairs (Fig.~\ref{fig:conv}) \cite{PhysRevA.69.032315, PhysRevResearch.4.013117}.

\begin{equation}
\begin{aligned}
U_{\text{conv}} 
&=\; \bigl[U_{3}(\theta_{1},\phi_{2},\lambda_{3}) \otimes U_{3}(\theta_{4},\phi_{5},\lambda_{6})\bigr] \\
&\quad\cdot \textsc{CNOT}_{1\rightarrow2} \cdot
      \bigl[R_{Y}(\theta_{7}) \otimes R_{Z}(\theta_{8})\bigr]                       \\
&\quad\cdot \textsc{CNOT}_{2\rightarrow1} \cdot
      \bigl[R_{Y}(\theta_{9}) \otimes \mathbb{I}\bigr]                             \\
&\quad\cdot \textsc{CNOT}_{1\rightarrow2} \cdot
      \bigl[U_{3}(\theta_{10},\phi_{11},\lambda_{12}) 
            \otimes U_{3}(\theta_{13},\phi_{14},\lambda_{15})\bigr].
\end{aligned}
\end{equation}

Here $U_{3}$ denotes the universal single-qubit rotation supplied by contemporary superconducting hardware,
\[
  U_{3}(\theta,\phi,\lambda)=R_{Z}(\phi)\,R_{Y}(\theta)\,R_{Z}(\lambda),
\]
and $\textsc{CNOT}_{a\rightarrow b}$ is a controlled-NOT with control on qubit~$a$ and target on qubit~$b$.  
Acting on a superposition, $U_{\text{conv}}$ extracts higher-order correlations that would require wide kernels in a classical CNN while consuming only two physical qubits.

\subsection{Quantum Pooling Layers}

\begin{figure*}[!ht]
    \centering
    \includegraphics[width=0.5\textwidth]{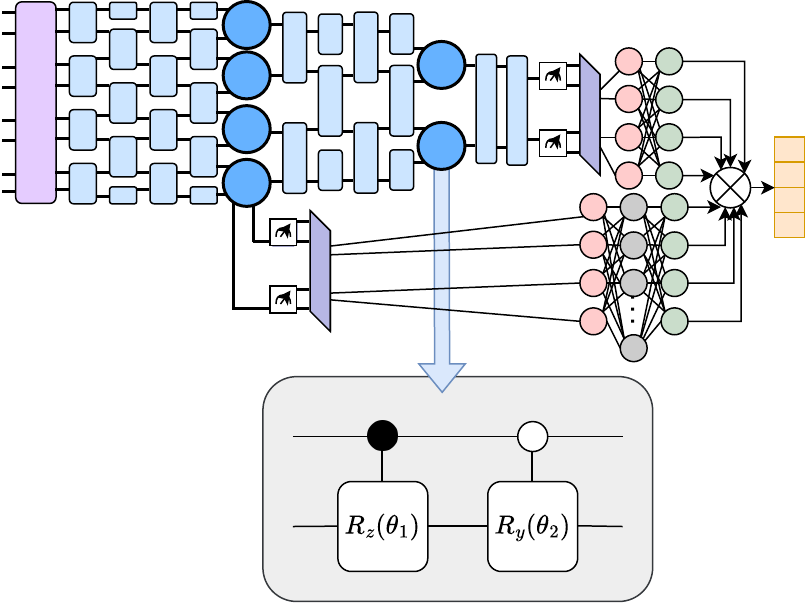}
    \caption[Quantum Pooling Operation]{
        Quantum Pooling Operation. This figure illustrates the application of a quantum pooling layer using a 2-parameter ansatz with controlled quantum gates (\( R_Z \) \& \( R_X \)). The pooling layer reduces qubit counts after the convolutional operations to manage computational complexity and mitigate the barren plateau problem by preserving the most informative qubits for further processing.
    }
    \label{fig:pooling}
\end{figure*}

A convolution is immediately followed by a two-qubit
\textit{quantum-pooling} gate $U_{\text{pool}}(\boldsymbol\phi)$ that discards
half of the register while preserving task-relevant correlations:
\begin{equation}
\begin{aligned} 
  U_{\text{pool}}
  \;=\;
  CR_Z(\phi_1)\,
  CR_X(\phi_2)
\end{aligned}
\end{equation}
where the control qubit survives and the target is traced out
(Fig.~\ref{fig:pooling}).  
This down-sampling (i) keeps gate count linear in the number of input pixels
and (ii) curbs barren-plateau gradients by limiting the Hilbert-space
dimension that remains in play \cite{PhysRevX.11.041011, Cong2019}.

\subsection{Layer-by-Layer Anatomy}
\begin{table*}[htbp]
  \centering
  \begin{tabular}{|l|l|c|c|}
    \hline
    \textbf{Stage} & \textbf{Operation} & \textbf{Qubit Count} & \textbf{Trainable Params} \\
    \hline
    Encoding    & Amplitude/ Angle map                                                     & $n$         & 0                   \\
    \textbf{Block 1} & QC 1 $\rightarrow$ QC 2 $\rightarrow$ Pooling             & $n/2$       & $2\times15 + 2$      \\
    \textbf{Block 2} & QC 3 $\rightarrow$ QC 4 $\rightarrow$ Pooling             & $n/4$       & $2\times15 + 2$      \\
    \textbf{Block 3} & QC 5 $\rightarrow$ QC 6                                   & $n/4$       & $2\times15$          \\
    Classical   & FC$_\text{ret}$ + FC$_\text{disc}$ $\rightarrow$ Ensemble $\rightarrow$ FC$_\text{final}$ & — & $\mathcal{O}(d\,c)$ \\
    \hline
  \end{tabular}
  \caption{End-to-End architecture showing quantum and classical stages.}
  \label{tab:end2end}
\end{table*}

The quantum backbone is organised into three macro-blocks: the first two follow a (conv, conv, pool) motif, while the third consists of a final (conv, conv) pair without pooling Fig.~\ref{fig:conv} and Fig.~\ref{fig:pooling}).

The quantum backbone is organised as three identical macro-blocks, each
comprising two convolutional stages followed by a pooling stage
(Fig.~\ref{fig:conv} and Fig.~\ref{fig:pooling}).  This
\texttt{(conv,\,conv,\,pool)} motif progressively compresses the register while
building increasingly abstract representations of the input.

\begin{enumerate}
  \item \emph{Local feature extraction.}  
        The first pair of convolutions acts on the angle-encoded image and
        detects short-range correlations; analogous to edge detectors in a
        classical CNN.  
        A subsequent pooling gate discards every second qubit, halving the
        register size.
  \item \emph{Intermediate abstraction.}  
        The reduced register is processed by the next two convolutional
        layers, which integrate information across a larger effective
        receptive field.
        A second pooling layer again halves the number of qubits, preserving
        only the qubits that carry the richest correlations.
  \item \emph{Global representation.}  
        The final convolutional pair operates on the remaining qubits to
        encode global patterns in the quantum state.
        No further pooling is applied, leaving the surviving qubits available
        for measurement and subsequent classical processing.
\end{enumerate}

\noindent Where, $n$ = initial qubits; $d$ = feature dimension after pooling; $c$ = number of classes.

\medskip

\noindent The joint quantum--classical loss
\begin{equation}
    \mathcal{L} = -\sum_{i=1}^{c} y_i \log \hat{y}_i
\end{equation}
is minimised via gradient back-propagation through the parameter-shift rule on the quantum side and standard autodiff on the classical side.
 
Once the final convolutional pair has acted, all surviving wires are measured in the computational basis.  We record two four-component probability vectors: one drawn from the qubits that remain in the data path after the first pooling layer and a second obtained from the qubits that were removed by pooling but are still physically present at the circuit’s end.  Each vector is linearly mapped from \([0,1]\) to the symmetric interval \([-2,2]\) using the affine rule \(p\rightarrow 4p-2\); this centring step greatly improves the conditioning of the
subsequent classical layers without introducing additional trainable parameters. The rescaled probabilities from the surviving wires pass through a single fully connected layer of width~\(c\) (the number of classes), while those from
the removed wires are processed by a lightweight two-layer head that briefly expands and then restores the feature dimension.  The two activations are combined element-wise, and the fused vector is fed to a final dense unit whose output constitutes the network logits.  Training minimises the cross-entropy
objective with gradients propagated end-to-end: standard automatic differentiation
updates all classical weights, and the same error signal continues through the
measurement operators via the parameter-shift rule to adjust every trainable
gate in the quantum circuit.  The entire hybrid therefore learns as a single,
coherent model.

\subsection{Quantum Pooling Mechanism}\label{ssec:qpool}

Quantum pooling plays the same down-sampling role as its classical analogue:
it prunes redundant degrees of freedom while preserving the features most
relevant to the task.  Each pooling unit is a compact two-gate sequence
\begin{equation}
  U_{\text{pool}}=CR_Z(\phi_1)\,CR_X(\phi_2),
\end{equation}
in which the control qubit is kept and the target qubit is measured and
removed (Fig.~\ref{fig:pooling}).  Because the two controlled rotations
entangle the pair immediately before the measurement, information about their
joint state is imprinted on the amplitudes of the surviving control qubit.
Consequently, correlations that would otherwise be lost remain accessible to
subsequent layers.

The design offers three practical benefits:
\begin{itemize}
  \item halving the register width keeps gate
        count and memory demands within NISQ limits;
  \item a smaller Hilbert space reduces the risk
        of barren-plateau gradients; and
  \item ater convolutions operate on a more
        informative subspace of the quantum state.
\end{itemize}

Inserting a pooling gate after every second convolution therefore yields the
quantum analogue of stride-2 pooling in classical CNNs, but implemented purely
with unitary operations.

\section{Proposed Utilization of Discarded Qubits and Classical Integration}\label{Proposed Utilization of Discarded Qubits and Classical Integration}

\subsection{Motivation and Significance}

Although quantum pooling layers systematically discard certain qubits, these discarded qubits may remain entangled or correlated, encapsulating valuable quantum information \cite{PhysRevA.98.062335}. Unlike conventional QCNN approaches where discarded qubits are irretrievably lost, the proposed architecture uniquely preserves these discarded qubit states for further classical processing. By retaining quantum correlations within the discarded qubits, the model enhances overall feature extraction and significantly improves classification accuracy.

By capturing both retained and discarded qubit states, the model reinforces feature extraction and classification through quantum correlations that otherwise would have gotten lost. This shows the full potential of a quantum system to allow the classical layers to process more subtle patterns from the input data.

\paragraph{Why only reuse the first set of discarded qubits.}
Although our backbone contains two pooling layers, we deliberately recycle only
the qubits discarded after the \emph{first} pooling operation. The second
pooling stage serves as the final dimensionality bottleneck, concentrating
information into a compact subsystem for hybrid readout \cite{Cong2019, PhysRevX.11.041011}. Reintroducing its
discarded qubits would counteract this compression and undermine the purpose of
pooling \cite{Cong2019}. Moreover, the qubits discarded at the final stage typically carry
residual entangled information that overlaps with the surviving ones, leading
to redundancy and amplified noise rather than genuinely new features \cite{larose2020robust, Caro_2023, McClean_2018}. In our
experiments, attempts to recycle both sets of discarded qubits yielded negligible or unstable gains, whereas reusing only the first set consistently
provided a stable improvement. We therefore fix our design choice to reuse the first discarded qubits only, striking a balance between richer feature utilisation and efficient compression at the quantum–classical interface.

\subsubsection{Proposed Scaling Method to Process Retained \& Discarded Qubit States}\label{ssec:scaling}

A pivotal part of the hybrid pipeline is a \emph{single} affine rescaling
that places quantum outputs in the numerical regime most amenable to classical
optimisation. Without this step the raw measurement probabilities
$p\in[0,1]$ would enter the classical layers highly skewed, hampering
gradient flow and obscuring the complementary value of the retained and
discarded streams.

\vspace{3pt}\noindent
Two deterministic checkpoints are defined:
\begin{itemize}
\item[\textbullet] \emph{Retained qubits} are measured after \emph{all}
quantum–convolution and pooling stages, i.e.\ at the logical end of
the quantum backbone.
\item[\textbullet] \emph{Discarded qubits} are measured immediately after
the first pooling layer, precisely when they leave the quantum data
path. This guarantees that the remaining gates act on a clean,
decoherence-free register.
\end{itemize}

\textbf{Why rescale?}
Classical activations such as $\tanh$ operate most
effectively on zero-centred, symmetric inputs with unit-scale variance. To
meet this criterion we map every probability through
\begin{equation}\label{eq:affine}
p_{\mathrm{scaled}} = 4p - 2, \qquad p\in[0,1],
p_{\mathrm{scaled}}\in[-2,2]
\end{equation}
The transformation is linear; hence information preserving and furnishes three
critical advantages:
\begin{enumerate}
\item \emph{Normalization.} Outputs are centred at the origin, aligning
with the most sensitive region of $\tanh$ and similar non-linearities.
\item \emph{Symmetry.} Positive and negative excursions are treated
equally, preventing biased weight updates during back-propagation.
\item \emph{Enhanced learning dynamics.} The four-fold expansion of the
raw range widens the gradient window, mitigating both vanishing and
exploding tendencies while preserving the relative ordering of
probabilities.
\end{enumerate}
The affine map in Eq.~(\ref{eq:affine}) therefore constitutes a simple yet
essential bridge: it preserves the full informational content of quantum 
measurements while placing that content squarely within the operational sweet spot of modern classical neural networks.
Applying the same rule to \emph{both} measurement
sites ensures that retained and discarded feature vectors occupy identical
numerical domains. As a result, subsequent dense layers can weigh the two
streams purely on the basis of learned relevance rather than arbitrary scale
differences. In practice this uniform treatment tightens the coupling between
quantum and classical components, accelerates convergence, and yields a more
robust decision surface all without adding a single trainable parameter.

\begin{enumerate}
\item \textbf{Preserving gradients under $\tanh$.}  
The derivative of $\tanh$ is largest for $|x|\!\lesssim\!2$. Setting $\kappa=4$ stretches the centred interval to $z\in[-2,2]$, ensuring that almost all activations lie within this high-sensitivity band. Smaller values of $\kappa$ leave activations clustered near the origin, where $\tanh$ behaves almost linearly, reducing expressive capacity. Larger values push activations into the saturation plateau ($|x|\gtrsim3$), where $\tanh'$ approaches zero and gradients vanish.

\item \textbf{Variance alignment with Glorot/Xavier initialisation.}  
If $p$ is approximately uniform, then $\operatorname{Var}[p]=1/12$. Scaling by $\kappa$ yields $\operatorname{Var}[z]=\kappa^{2}/12$. The Glorot criterion suggests maintaining variance close to one for stable forward and backward signal propagation, which implies $\kappa\approx\sqrt{12}\simeq3.46$ \cite{pmlr-v9-glorot10a}. Choosing $\kappa=4$ provides a simple round number, slightly over-spreading the activations (helpful for $\tanh$) while remaining below the saturation threshold.

\item \textbf{Hardware-friendly dynamic range.}  
Many fixed-point inference engines (e.g., low-power edge accelerators intended to host the classical head alongside a NISQ device) encode activations in signed 8-bit or 16-bit formats with two integer bits. The interval $[-2,2]$ maps neatly into this representable range, so $\kappa=4$ avoids additional rescaling or clipping during deployment.
\end{enumerate}

To avoid introducing an additional hyper-parameter and to ensure reproducibility, we fix $\kappa=4$ in all experiments. This choice is supported by gradient analysis (keeping $z\in[-2,2]$ in the high-slope, non-saturating region of $\tanh$), variance considerations ($\operatorname{Var}(z)=\kappa^2/12\approx1.33$ at $\kappa=4$, compatible with Glorot/Xavier), and practical deployment ranges.

\subsection{Proposed Classical Neural Network Layers to Process Extracted Features} 

The output states; specifically, the rescaled expectation values obtained from measuring the retained and discarded qubits (equivalently, their corresponding probabilities mapped to the interval $[-2,2]$), these are then fed into a series of classical fully connected layers. These are critical in the hybrid quantum-classical architecture for refining the features extracted by those quantum means and fusing them with classical computations to yield a final classification output. It ensures that each classical layer involved is crafted with the appropriate trade-off between model complexity and computational efficiency to manage quantum data effectively.

\subsubsection{Processing the Retained-Qubit Features}
\label{sssec:ret-path}

The $m$ measurement outcomes that survive the quantum-convolution and pooling stages form a compact, high-order feature vector. We pass this vector through a \emph{single} fully connected layer of width $m$ matching the number of target classes followed by a $\tanh$ nonlinearity.

Let
\[
  \mathbf{x}_{\mathrm{ret}} \;\in\; [-2,2]^m
\]
be the rescaled measurement vector (see Sec.~\ref{ssec:scaling}).  The retained-qubit pathway computes
\begin{align}
  \widetilde{\mathbf{y}}_{\mathrm{ret}}
  = \tanh\bigl(W_{\mathrm{ret}}\,\mathbf{x}_{\mathrm{ret}} + \mathbf{b}_{\mathrm{ret}}\bigr),\\
  W_{\mathrm{ret}}\in\mathbb{R}^{m\times m},\;\;
  \mathbf{b}_{\mathrm{ret}}\in\mathbb{R}^m,\nonumber
\end{align}
yielding the refined feature vector $\widetilde{\mathbf{y}}_{\mathrm{ret}}\in\mathbb{R}^{m}$.

This design offers two practical advantages:
\begin{enumerate}

    \item \textbf{Parameter efficiency:} the retained-qubit head contains exactly
        $m^{2}$ weights and $m$ biases.  In the four-class experiments
        reported here ($m=4$) this is only $4^{2}+4=20$ trainable parameters.
        By contrast, even a minimal classical head that first flattens a
        $128$-feature tensor and maps it to four logits would require
        $128\times4+4=516$ parameters, and standard CNN classifiers often
        employ $10^{4}$–$10^{6}$ weights.  Thus the proposed layer is one to
        two orders of magnitude lighter while still providing a non-linear
        transform. 
  \item \textbf{Seamless Fusion:} $\widetilde{\mathbf{y}}_{\mathrm{ret}}$ shares both dimensionality and scaling with the discarded-qubit branch (Sec.~\ref{sssec:disc-path}), allowing direct dot-product fusion without further projection.
\end{enumerate}

By keeping the classical head shallow and well-conditioned, the network preserves quantum-derived information while minimising overhead and facilitating stable end-to-end optimisation.

\subsubsection{Processing the Discarded‐Qubit Features}
\label{sssec:disc-path}

Although pooling removes half the qubits, their measurements remain informative via prior entanglement. We measure the discarded register, rescale the outcomes to $[-2,2]$, and obtain
\[
  \mathbf{x}_{\mathrm{disc}}\in[-2,2]^m.
\]
These values are then fed into a three‐stage dense subnetwork that mirrors the retained path in width but enhances expressivity:

\begin{enumerate}
  \item \textbf{Class‐Aligned Projection:}
  \begin{align*}
      \mathbf{y}_{\mathrm{disc}}
      = \tanh\bigl(W_{\mathrm{disc}}\,\mathbf{x}_{\mathrm{disc}} + \mathbf{b}_{\mathrm{disc}}\bigr)\\
      W_{\mathrm{disc}}\in\mathbb{R}^{m\times m},\;
      \mathbf{b}_{\mathrm{disc}}\in\mathbb{R}^m.
  \end{align*}

  \item \textbf{Deep Feature Expansion:}\\
    To uncover higher‐order correlations, we expand the feature dimension by factor $k$:
    \begin{align*}
        \mathbf{h}_{\mathrm{disc}}
      = \tanh\bigl(W_{h}\,\mathbf{y}_{\mathrm{disc}} + \mathbf{b}_{h}\bigr),\\
      W_{h}\in\mathbb{R}^{mk\times m},\;
      \mathbf{b}_{h}\in\mathbb{R}^{mk}.
    \end{align*}

  \item \textbf{Dimensionality Recovery:}\\
    A final contraction restores the original width:
    \begin{align*}
        \widetilde{\mathbf{y}}_{\mathrm{disc}}
      = \tanh\bigl(W_{\mathrm{out}}\,\mathbf{h}_{\mathrm{disc}} + \mathbf{b}_{\mathrm{out}}\bigr),\\
      W_{\mathrm{out}}\in\mathbb{R}^{m\times mk},\;
      \mathbf{b}_{\mathrm{out}}\in\mathbb{R}^m.
    \end{align*}
\end{enumerate}

This design injects only $mk + m$ classical neurons that are lightweight compared to typical CNN heads while ensuring both quantum‐derived streams share identical dimensionality and scaling for seamless dot‐product fusion.

\subsubsection{Proposed Final Integration and Classification}
\label{sssec:fusion}

After their independent refinements (Secs.~\ref{sssec:ret-path}--\ref{sssec:disc-path}), the two quantum‐derived feature streams are merged \emph{without} adding any further trainable parameters. Denote
\[
  \mathbf{y}_{\mathrm{ret}} = \bigl(y_{\mathrm{ret},1}, \ldots, y_{\mathrm{ret},m}\bigr), 
  \quad
  \mathbf{y}_{\mathrm{disc}} = \bigl(y_{\mathrm{disc},1}, \ldots, y_{\mathrm{disc},m}\bigr)
\]
the outputs of the retained‐ and discarded‐qubit pathways. Fusion is performed via the Hadamard (element‐wise) product:
\begin{equation}
  \mathbf{z}
  = \mathbf{y}_{\mathrm{ret}} \odot \mathbf{y}_{\mathrm{disc}}
  = \bigl(y_{\mathrm{ret},1}y_{\mathrm{disc},1}, \dots, y_{\mathrm{ret},m}y_{\mathrm{disc},m}\bigr).
  \label{eq:hadamard}
\end{equation}
For the benchmark ($m=4$), this reduces to
\begin{equation}
\mathbf{z}
= \begin{pmatrix}
  y_{\mathrm{ret},1}\cdot y_{\mathrm{disc},1} \\[6pt]
  y_{\mathrm{ret},2}\cdot y_{\mathrm{disc},2} \\[6pt]
  y_{\mathrm{ret},3}\cdot y_{\mathrm{disc},3} \\[6pt]
  y_{\mathrm{ret},4}\cdot y_{\mathrm{disc},4}
\end{pmatrix}.
\end{equation}

\paragraph{Hadamard Product}
\begin{itemize}
  \item \textbf{Correlation:} multiplicative terms $y_{\mathrm{ret},i}y_{\mathrm{disc},i}$ capture class‐aligned interactions unavailable to either branch alone.
  \item \textbf{Parameter economy:} the fusion is strictly algebraic, introducing \emph{no} extra weights.
  \item \textbf{Symmetry:} the commutativity of $\odot$ ensures both streams contribute equally.
\end{itemize}

\paragraph{Why choose the Hadamard (element-wise) product?}
Multiplication behaves as a soft logical \textit{AND}: a fused entry
\(z_i = y^{\text{ret}}_i\,y^{\text{disc}}_i\) becomes large \emph{only when
both} the retained- and discarded-qubit branches agree on class~\(i\).  This
gating effect encourages the two heads to learn complementary features and
suppresses spurious confidence from either branch in isolation.  Simple
addition, in contrast, acts like a soft \textit{OR}; one highly confident
branch can dominate the sum, diluting the benefit of quantum-classical
agreement.  The Hadamard product also keeps the activations in the same
\([-1,1]\) range as the inputs, preserving the variance that Glorot
initialisation expects and avoiding extra normalisation layers or tuning \cite{pmlr-v9-glorot10a}.
Similar multiplicative fusion appears in FiLM conditioning, attention gating
and GRU/LSTM \cite{DBLP:journals/corr/ChungGCB14, hochreiter1997long} update gates, where agreement between pathways is crucial.  Our
preliminary trials showed that addition or concatenation offered no
performance gain but required more parameters or hyper-parameter tuning,
whereas the parameter-free Hadamard product delivered the sharpest decision
boundaries while keeping the model ultra-lightweight.

\paragraph{Direct optimisation}
The fused vector $\mathbf{z}$ is treated as the \emph{logit} output of the
network.  Training minimises the categorical cross-entropy over a mini-batch of
$N$ samples,
\begin{equation}
  \mathcal{L}
  \;=\;
  -\frac{1}{N}\sum_{n=1}^{N}\;
  \sum_{i=1}^{c} \;y^{(n)}_i \,\log \hat{y}^{(n)}_i ,
  \label{eq:xent_fused}
\end{equation}
where $\mathbf{y}^{(n)}$ is the one‐hot ground‐truth label. No auxiliary dense layers, extra $\tanh$, softmax, or sigmoid are used during training; gradients flow unobstructed through the Hadamard product back into both dense heads and via the parameter‐shift rule into the quantum circuit. If calibrated probabilities are required at inference, a softmax can be applied \emph{post‐training} without altering learned parameters.

\paragraph{End-to-end back‐propagation}
Because the Hadamard product is differentiable, the same loss drives a seamless gradient pipeline:
\begin{align*}
    \nabla_z \mathcal{L}  &\;\longrightarrow\; \bigl(\nabla_{\mathbf{y}_{\mathrm{ret}}}\mathcal{L},\;\nabla_{\mathbf{y}_{\mathrm{disc}}}\mathcal{L}\bigr)\\  &\;\longrightarrow\;  \nabla_{\mathbf{p}}\mathcal{L}\\  &\;\xrightarrow{\;\text{parameter‐shift}\;}\;
  \nabla_{\boldsymbol{\theta}}\mathcal{L}\,.
\end{align*}

\begin{itemize}
  \item The error signal first traverses the Hadamard product, distributing sensitivity to both dense branches.
  \item Standard automatic differentiation updates all classical weights.
  \item The resulting measurement gradients are converted, via the parameter‐shift rule, into updates for every trainable gate in the quantum circuit.
\end{itemize}
Thus, back‐propagation is truly end‐to‐end; the hybrid network is optimised as a single unit, with classical and quantum parameters adapting cooperatively to minimise a unified objective.

\section{Evaluation of Proposed Hybrid Quantum-Classical Model}\label{Evaluation of Proposed Hybrid Quantum-Classical Model}

We evaluated our hybrid quantum–classical QCNN on three benchmark image-classification datasets: MNIST, Fashion-MNIST, and OrganAMNIST (a MedMNIST subset) \cite{medmnistv2}; to quantify the effect of its key design choices. Two variants were therefore benchmarked: a \textit{recycled branch} that integrates both retained- and discarded-qubit measurements, and a \textit{baseline} that omits the latter while keeping all other settings identical. Evaluation followed standard practice, reporting top-1 accuracy, F1-score, precision, and recall, and was supplemented by loss/accuracy learning curves and class-normalised confusion matrices.

\begin{table*}[htbp]
  \centering
  \begin{tabular}{@{} l l l l c c c c c @{}}
    \toprule
    \textbf{Dataset} & \textbf{Class Split} & \textbf{Encoding} & \textbf{Qubits} 
    & \textbf{Accuracy} & \textbf{F1} & \textbf{Recall} & \textbf{Precision} \\
    \midrule
    MNIST & 0,1,2,3 & angle     & 7,1 & 88.55 & 88.39 & 88.55 & 88.39 \\
          &         & amplitude & 7,1 & \textbf{93.55} & 93.53 & 93.55 & 93.53 \\
          &         & angle     & 3,5 & 89.67 & 89.75 & 89.75 & 89.71 \\
          &         & amplitude & 3,5 & 90.15 & 89.45 & 89.35 & 89.8  \\
          & 3,4,5,6 & angle     & 7,1 & 70.32 & 69.24 & 70.43 & 70.00 \\
          &         & amplitude & 7,1 & \textbf{88.52} & \textbf{88.48} & \textbf{88.52} & \textbf{88.52} \\
          &         & angle     & 3,5 & 87.71 & 87.64 & 87.71 & 87.94 \\
          &         & amplitude & 3,5 & 88.25 & 88.30 & 88.35 & 88.35 \\
    Fashion-MNIST & 0,1,2,3 & angle     & 7,1 & 68.75 & 68.65 & 68.75 & 68.86 \\
                  &         & amplitude & 7,1 & \textbf{86.55} & \textbf{86.53} & \textbf{86.55} & \textbf{86.58} \\
                  &         & angle     & 3,5 & 67.68 & 67.36 & 67.68 & 68.35 \\
                  &         & amplitude & 3,5 & 85.93 & 85.03 & 85.93 & 85.39 \\
                  & 1,2,8,9 & angle     & 7,1 & 79.9  & 79.1  & 79.95 & 79.33 \\
                  &         & amplitude & 7,1 & 95.25 & 95.29 & 95.25 & 95.43 \\
                  &         & angle     & 3,5 & 80.36 & 80.16 & 80.36 & 80.35 \\
                  &         & amplitude & 3,5 & \textbf{96.45} & \textbf{96.46} & \textbf{96.45} & \textbf{96.55} \\
    OrganAMNIST   & 0,3,7,8 & amplitude & 7,1 & 82.59 & 82.44 & 82.59 & 83.64 \\
                  &         & amplitude & 3,5 & \textbf{88.5}  & \textbf{88.46 }& \textbf{88.5}  & \textbf{88.5}  \\
    \bottomrule
  \end{tabular}
  \caption{Detailed performance metrics for each dataset, class split, encoding method, and qubit configuration. Metrics shown are accuracy, F1 score, recall, and precision (all in \%).}
  \label{tab:full_results}
\end{table*}

\begin{table*}[htbp]
\centering
\begin{tabular}{lllccccc}
\toprule
\textbf{Dataset} & \textbf{Class Split} & \textbf{Encoding} & \textbf{Accuracy (\%)} & \textbf{F1 Score (\%)} & \textbf{Recall (\%)} & \textbf{Precision (\%)} \\
\midrule
MNIST & 0,1,2,3   & angle     & 71.85 & 71.49 & 71.85 & 71.73 \\
 &   & amplitude & 70.03 & 67.92 & 70.03 & 71.16 \\
 & 3,4,5,6   & angle     & 60.39 & 52.94 & 60.39 & 57.78 \\
 &   & amplitude & 68.51 & 67.49 & 68.51 & 68.58 \\
Fashion-MNIST & 0,1,2,3 & angle     & 59.2  & 52.08 & 59.2  & 48.05 \\
 &  & amplitude & 74.3  & 73.99 & 74.79 & 74.79 \\
 & 1,2,8,9 & angle     & 60.4  & 55.53 & 60.4  & 64.62 \\
 &  & amplitude & 81.4  & 81.48 & 81.23 & 81.41 \\
OrganAMNIST   & 0,3,7,8 & amplitude & 69.13 & 64.36 & 69.13 & 61.41 \\
\bottomrule
\end{tabular}
\caption{Test performance on MNIST, Fashion-MNIST, and OrganAMNIST \textit{without} discarded qubit reuse.}
\label{tab:no_discarded_qubits}
\end{table*}

\begin{figure*}[ht!]
\centering
\includegraphics[width=1\textwidth]{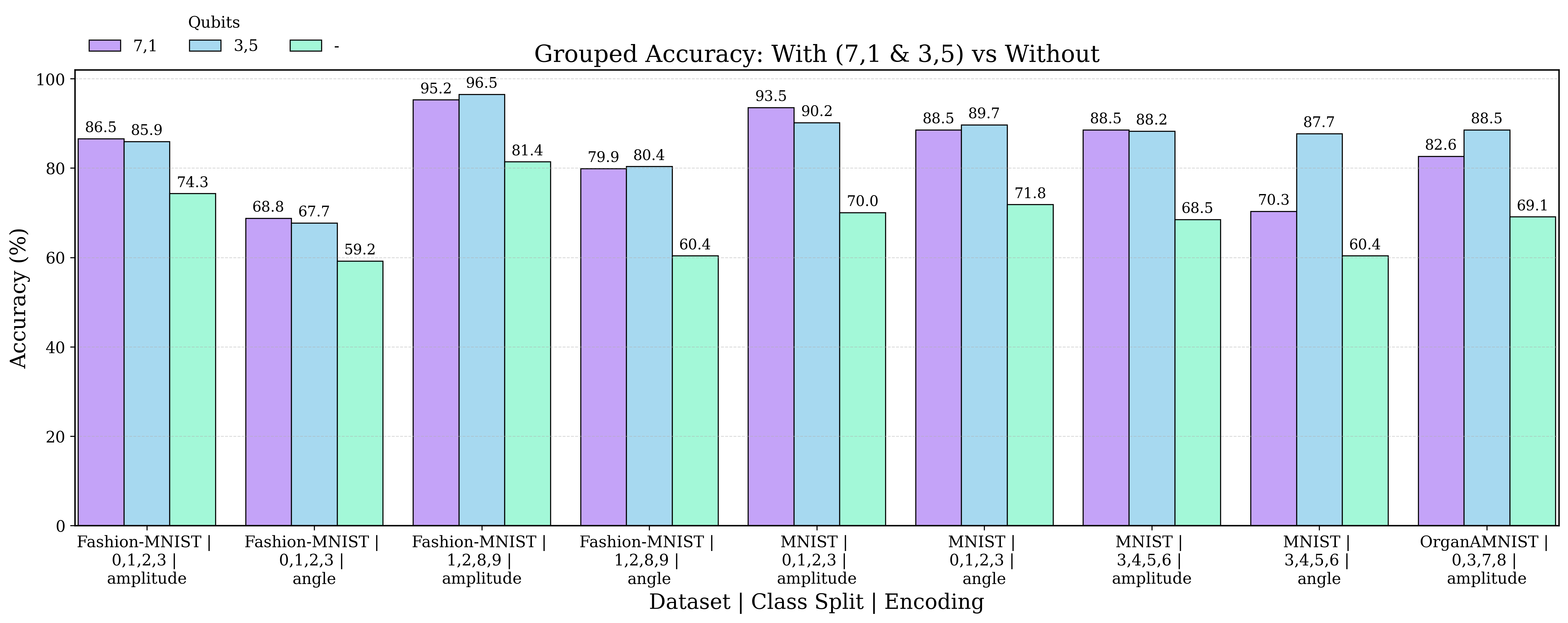}
\caption{Grouped accuracy comparison across datasets, class splits, and encoding types, contrasting configurations that reuse discarded qubits and without those.}
\label{Grouped accuracy}
\end{figure*}

Incorporating the discarded qubits yielded pronounced gains: test accuracy climbed from 68.51\% to 93.55\% on MNIST, from 81.40\% to 96.45\% on Fashion-MNIST, and from 69.13\% to 88.50\% on OrganAMNIST, with commensurate improvements in F1 ($\approx$ +21–37 percentage points), precision, and recall. The learning curves in Figure \ref{fig:fmnist_acc_combo} show faster convergence and a consistently smaller train–test gap for the recycled branch, indicating improved generalisation. When set beside recent quantum and hybrid baselines, namely the QCNN of Bokhan et al.,\cite{bokhan2022multiclass} and the QMCC of Chalumuri et al.,\cite{Chalumuri}, our model achieves the highest reported accuracies while keeping a compact 262-parameter footprint.

In sum, the results confirm that, when paired with the proposed scaling and joint-optimisation strategy, utilizing information from qubits otherwise discarded at the pooling stage consistently and substantially elevates classification performance across diverse image domains.

\subsection{Quantitative Performance Summary}

Table~\ref{tab:full_results} reports the full test-set metrics- accuracy, F1-score, recall and precision; for every dataset, class split, encoding scheme and qubit arrangement when discarded-qubit measurements are recycled, while Table~\ref{tab:no_discarded_qubits} lists the corresponding results when those measurements are omitted. Taken together, the two tables provide a unified view of the quantitative benefit conferred by the proposed qubit-reuse strategy.

The baseline keeps the quantum circuit exactly as-is up to each pooling operation. At that point, however, it follows the conventional QCNN pattern:both qubits in a two-qubit pooling pair are measured, one wire is discarded, and no information from those measurements is forwarded to later layers. The classical post-processing stack is therefore just a single fully connected layer tied to the retained qubits. All architectural and training hyper-parameters as well as state-encoding scheme, circuit depth (six convolutional blocks plus three pooling stages), optimiser, batch size, \& number of training steps are kept identical so that the only difference is the presence or absence of the recycled branch. Across all three benchmarks the pattern is unequivocal. Recycling discarded qubits raises accuracy by 22-28 percentage points and yields commensurate improvements in F1, precision, and recall. On the MNIST 0-1-2-3 split, amplitude encoding with a (7,1) qubit pair (the probabilities from retained wire 7 \& 1) attains 93.3\% accuracy versus 70.0\% for the baseline (Table~\ref{tab:full_results} vs Table~\ref{tab:no_discarded_qubits}). Fashion-MNIST shows a similar gain: accuracy rises from 81.4\% to 96.45\% , with amplitude encoding again providing the largest boost. The effect is most dramatic on the class-imbalanced OrganAMNIST subset, where accuracy leaps from 69.13\% to 88.5\%  and the F1-score nearly doubles.

A clear encoding trend also emerges. Amplitude embeddings consistently outperform angle embeddings under otherwise identical settings, suggesting that the global amplitude map exploits the added information from discarded qubits more effectively. Angle-based embeddings also gain from qubit reuse; for example, accuracy on the MNIST 3-4-5-6 split climbs from 60.39\% to 88.52 \% which yet they still fall short of their amplitude-encoded counterparts.

For rapid visual comparison we recommend a grouped bar chart (Figure~\ref{Grouped accuracy}) that plots accuracy and F1 for each dataset under both configurations, and a heat map of $\Delta$-accuracy that highlights per-class improvements. Both graphics make the scale and consistency of the gains immediately apparent and complement the numerical detail in Tables~\ref{tab:full_results} and~\ref{tab:no_discarded_qubits}.

\begin{figure}[htbp]
    \centering
    \begin{subfigure}{0.5\textwidth}
        \centering
        \includegraphics[width=\linewidth]{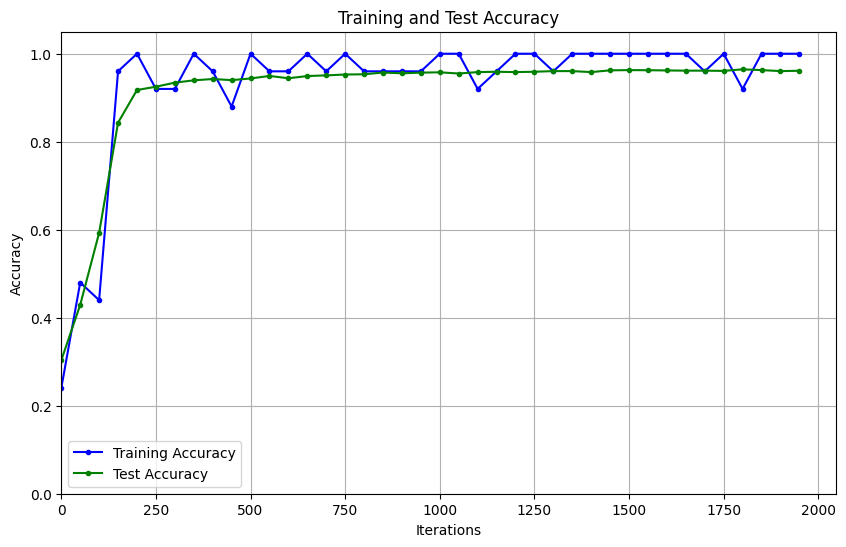}
        \caption{With discarded-qubit reuse. Accuracy climbs rapidly and the training-vs-test traces remain tightly coupled.}
    \end{subfigure}
    \hfill
    \begin{subfigure}{0.5\textwidth}
        \centering
        \includegraphics[width=\linewidth]{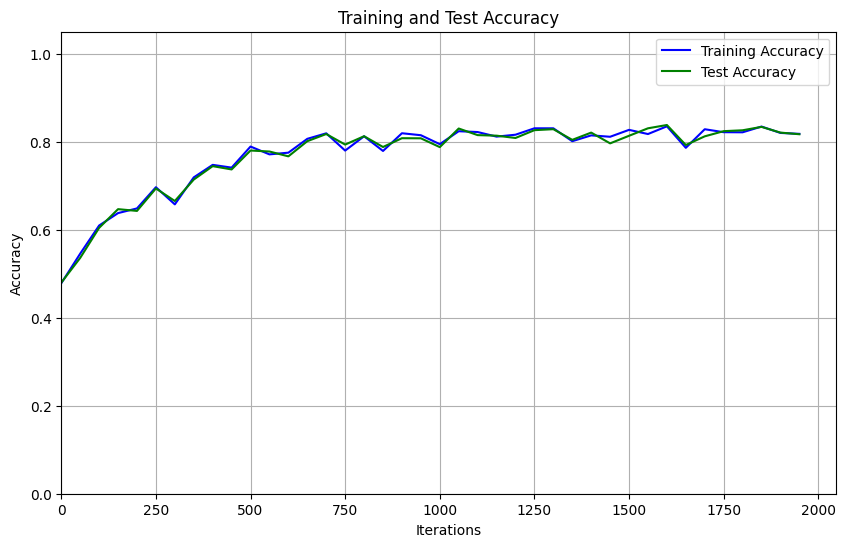}
        \caption{Without reuse (baseline). Accuracy levels off early and a wide gap opens between training and test curves.}
    \end{subfigure}
    \caption{Fashion-MNIST 1-2-8-9; training/test accuracy comparison. Using measurements from qubits normally discarded by pooling yields both higher accuracy and a much smaller generalisation gap.}
    \label{fig:fmnist_acc_combo}
\end{figure}

Figure~\ref{fig:fmnist_acc_combo} contrasts the complete training-versus-test accuracy traces for the recycled and baseline QCNNs on the most challenging Fashion-MNIST split. Reusing discarded qubits propels the network to \textbf{95--96\%} test accuracy after roughly 600 iterations, while the baseline saturates at approximately \textbf{81\%}. The recycled model also narrows the generalisation gap to less than \textbf{3 percentage points} throughout training; in contrast, the baseline’s training curve flattens after ~450 iterations; from that point onward the generalisation gap widens to more than 10 percentage points.

\begin{figure}[htbp]
    \centering
    \begin{subfigure}{0.5\textwidth}
        \centering
        \includegraphics[width=\linewidth]{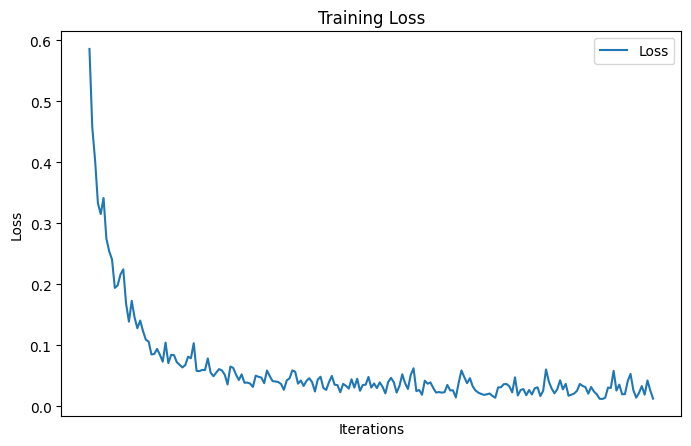}
        \caption{With discarded-qubit reuse. Loss falls steeply at the start and settles into a smooth, low plateau.}
    \end{subfigure}
    \hfill
    \begin{subfigure}{0.5\textwidth}
        \centering
        \includegraphics[width=\linewidth]{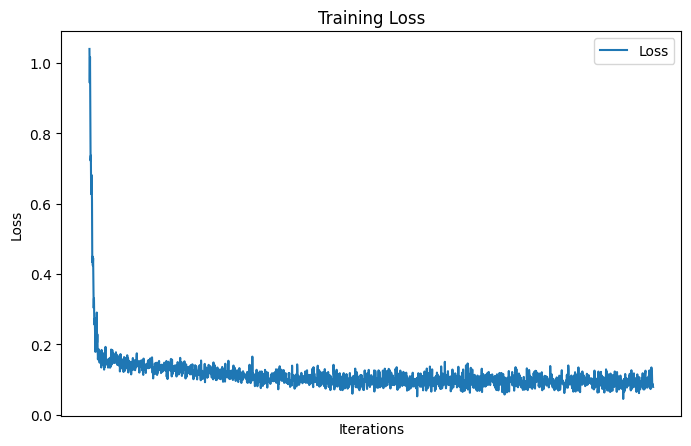}
        \caption{Without reuse (baseline). Loss declines more slowly, stabilising at a higher, noisier level.}
    \end{subfigure}
    \caption{Fashion-MNIST 1-2-8-9; training-loss comparison. The recycled-qubit branch drives faster optimisation and converges to a cleaner minimum than the baseline.}
    \label{fig:fmnist_loss_combo}
\end{figure}

Figure \ref{fig:fmnist_loss_combo} tells the same story. When discarded qubits are recycled, the objective drops by an order of magnitude within the first 100 iterations and then plateaus below \textbf{0.07}. Omitting those measurements leaves the loss stuck around \textbf{0.15} and noticeably noisier, a clear sign of slower convergence and a less reliable optimum.

Taken together, the curves confirm that the improvements come from genuinely faster optimisation and stronger generalisation, not from over-fitting or post-hoc calibration. The recycled branch drops steeply because feeding discarded-qubit statistics back into the network gives the optimiser a richer gradient signal. With more information at each step, the search moves decisively toward a favourable region of parameter space, after which the loss declines smoothly with little noise. The baseline, in contrast, must learn from a thinner and noisier feature set; its gradients vanish quickly, the objective stalls, and the remaining fluctuations are little more than stochastic noise. A similar, though slightly milder, pattern holds for every other dataset and class split we evaluated.


\subsection{Comparison to Prior Work}
\label{sec:comparison-prior-work}

Table~\ref{tab:comparison-simple} benchmarks our 262-parameter hybrid QCNN against two of the most parameter-frugal baselines in the literature: (i) the quantum convolutional neural network (QCNN) of Bokhan \emph{et al.} \cite{bokhan2022multiclass}, and  (ii) the quantum multi-class classifier (QMCC) proposed by Chalumuri  \emph{et al.} \cite{Chalumuri}. Both baselines deliberately keep circuit depth and trainable parameters low in order to remain hardware-friendly. Across every MNIST and Fashion-MNIST split our network achieves the highest accuracy, edging past Bokhan’s QCNN and leaving the QMCC of Chalumuri \emph{et al.} further behind.
Crucially, these gains do not rely on deeper circuits, extra ancilla qubits, or added classical post-processing. The sole architectural change is the novel recycled-qubit branch itself: its measurements are passed through one affine-rescaling layer and then incorporated into the same joint quantum–classical optimisation used for the rest of the model.

\begin{table}[t]
  \centering
  \renewcommand{\arraystretch}{1.15}
  \setlength{\tabcolsep}{4pt} 
  \begin{tabularx}{\columnwidth}{@{} l *{3}{>{\centering\arraybackslash}X} @{}}
    \toprule
    \textbf{Dataset} & \textbf{Chalumuri} & \textbf{Bokhan} & \textbf{Proposed} \\
    \addlinespace[-2pt]
                     & \scriptsize(–)~\cite{Chalumuri} 
                     & \scriptsize(188 Params)~\cite{bokhan2022multiclass} 
                     & \scriptsize(262 Params) \\
    \midrule
    MNIST (3456)          & 71.44 & 85.14 & \textbf{88.52} \\
    MNIST (0123)          & 77.64 & 90.03 & \textbf{93.55} \\
    Fashion-MNIST (0123)  & 71.15 & 85.93 & \textbf{86.55} \\
    Fashion-MNIST (1289)  & 79.33 & 93.63 & \textbf{96.45} \\
    \bottomrule
  \end{tabularx}
  \caption{Comparison of lightweight quantum models on four-class splits. Parameter counts are shown in parentheses under each method’s name. The last column is the test accuracy of the proposed model.}
  \label{tab:comparison-simple}
\end{table}

The recycled-qubit branch lifts accuracy beyond any modification previously reported for compact QCNNs, while adding only 74 trainable weights relative to Bokhan’s architecture. In other words, the proposal delivers a new state-of-the-art while preserving the ultra-light hardware profile that makes QCNNs attractive for current NISQ devices; every performance gain can be traced directly to information that earlier networks were discarding.

\section{Conclusion}\label{Conclusion}
This work proposes a new quantum-classical hybrid architecture that exploits the discarded qubits in the conventional QCNN structure along with joint optimization of the classical classifier \& quantum network. Our work shows that the re-inclusion of qubits that had previously been discarded in the literature actually retains valuable quantum information that is capable of yielding better prediction performance and more accuracy in such datasets as MNIST, Fashion MNIST, and OrganAMNIST. Results indicate that misclassifications in the proposed architecture are lower compared to those with no discarded qubits and other QCNN models. This work further develops a hybrid quantum-classical architecture that employs to great advantage the best of quantum and classical processing, contributing this way to robustness and accuracy in machine learning models. The results highlight how the proposed architecture produces very satisfactory results on the conducted experiment.



\bibliography{references}



\end{document}